\documentclass[conference]{IEEEtran}
\IEEEoverridecommandlockouts
\usepackage{cite}
\usepackage{amsmath,amssymb,amsfonts}
\usepackage{algorithmic}
\usepackage{graphicx}
\usepackage{textcomp}
\usepackage{xcolor}
\usepackage{multirow}
\def\BibTeX{{\rm B\kern-.05em{\sc i\kern-.025em b}\kern-.08em
    T\kern-.1667em\lower.7ex\hbox{E}\kern-.125emX}}
\begin{document}


\title{Numerical Comparisons of Linear Power Flow Approximations: Optimality, Feasibility, and Computation Time
}

\author{
\IEEEauthorblockN{Meiyi Li }
\IEEEauthorblockA{\textit{Carnegie Mellon University}\\
Pittsburgh, PA, USA\\
meiyil@andrew.cmu.edu
}

\and
\IEEEauthorblockN{Yuhan Du, Javad Mohammadi}
\IEEEauthorblockA{\textit{University of Texas Austin}\\
Austin, TX, USA\\
\{yuhandu,\\ javadm\}@utexas.edu}

\and
\IEEEauthorblockN{ Constance Crozier, Kyri Baker}
\IEEEauthorblockA{\textit{University of Colorado Boulder}\\
Boulder, CO, USA\\
\{constance.crozier,\\ kyri.baker\}@colorado.edu}
\and
\IEEEauthorblockN{Soummya Kar}
\IEEEauthorblockA{\textit{Carnegie Mellon University}\\
Pittsburgh, PA, USA\\
soummyak@andrew.cmu.edu
}

}

\maketitle

\begin{abstract}
Linear approximations of the AC power flow equations are of great significance for the computational efficiency of large-scale optimal power flow (OPF) problems. Put differently, the feasibility of the obtained solution is essential for practical use cases of OPF. However, most studies focus on approximation error and come short of comprehensively studying the AC feasibility of different linear approximations of power flow. This paper discusses the merits of widely-used linear approximations of active power in OPF problems. The advantages and disadvantages of the linearized models are discussed with respect to four criteria; accuracy of the linear approximation, optimality, feasibility, and computation time. Each method is tested on five different systems. 
\end{abstract}


\section{Introduction}

Linear approximations of the AC power flow equations play an essential role in the modern electricity grid, in the face of new challenges in power grids. In the California ISO electricity market, the locational marginal prices need to be broadcast every 15 minutes\cite{intro1}, meaning that these variables need to be calculated across a large system on a relatively fast timescale. Additionally, the higher penetration of distributed energy resources makes the power network even more complicated\cite{intro2}. Although AC optimal power flow is considered the most physically representative, intrinsic non-linearity makes it unsuitable for rapid calculations. Instead, a method that balances the accuracy and efficiency would be preferable. Because of these factors, linear approximations prevail in applications like contingency selection, 
state estimation,
and calculation of locational marginal prices in electricity markets.

Many linear models have been developed for power flow. One of the most canonical models is the widely adopted DC power flow model. This linear approximation relies on engineering observations and is relatively simple to implement while sacrificing voltage magnitudes and line power losses\cite{review2}. To overcome these two downsides, subsequent models have been developed in the literature. Reference\cite{intro6} introduced a concept of network losses equivalent power to raise the accuracy of active power flow. To get a more precise voltage magnitude, Reference\cite{intro7} decoupled and linearized power variables. Meanwhile, linear approximated optimal power flow has been adopted in more new scenarios, confronting new challenges in power grids. Reference\cite{intro8} proposed an approximate linear three-phase power flow model in active distribution networks with many PV nodes. Moreover, Reference\cite{intro2} adopted a linearized optimal power flow and loss factors to build a novel distribution locational marginal price model. These advancements are promising for many future applications.

In all of the above linear approximations, there are many assumptions used to simplify the model, and hence the obtained solution may not be AC feasible, possibly resulting in unrealizable generation dispatch decisions \cite{BakerACM} or voltage violations in practice. Reference \cite{feasibility1} presented a unified DC power flow method to check the real power flow feasibility by finding feasible solutions of a set of linear inequalities. Reference \cite{feasibility2} further analyzes the feasibility of the linear approximation of both real and reactive power flows. However, these studies are based on the proportional relationship between real power and voltage phase angles and the proportional relationship between reactive power and voltage magnitudes. There has not been further feasibility research of other linear approximations of power flow. 

Several review papers for linear power flow models have been published. Authors in \cite{kargarian2016toward} compared different decomposition methods to speed up the convergence process of DC OPF. Reference \cite{review1} compares three decoupled linearized equivalent power flow
models and then suggests their scope of application. A theoretical analysis of key techniques used in different linear approximations is presented in \cite{review2}. The feasibility of the obtained solution is essential for practical applications, but detailed analyses of linear approximations of power flow are lacking in the existing literature. This paper aims to cover this gap. We provide a computational investigation across multiple test networks of how different linear approximations perform with respect to the accuracy, optimality, feasibility, and calculation time.

This paper is organized as follows:
Section II introduces five linearization methods for the AC power flow equations and further introduces two iterative methods based on linear power flow considering line loss. The accuracy, optimality, feasibility, and running time of these seven linear methods are presented in section III. Finally, section IV presents a superiority evaluation for different methods. 
\vspace{-0.04in}
\section{Linear Power Flow Approximations}
 The classical AC power balance ensures that supply matches demand given network parameters. The active power balance portion of the AC power flow equations can be written as:
\begin{align}
    &P_{i}^{G}-P_{i}^{L}=\sum_{j\in i}P_{ij}\label{bus}\\
    P_{ij}=g_{ij} (  V_{i}^{2}&-V_{i}V_{j}\cos(\theta _{i}-\theta _{j}) )-b_{ij}V_{i}V_{j}\sin(\theta _{i}-\theta _{j}))\label{acpf}
\end{align}

Here, $P_{i}^{G}$ and $P_{i}^{L}$ denote the active power generation and load of bus $i$, respectively. $V_{i}$ and $\theta_{i}$ are the voltage magnitude and the voltage phase, respectively. Line parameters $g_{ij}$ and $b_{ij}$ are the conductance and the susceptance of line ($i$, $j$). Lastly, $P_{ij}$ is the active power flow on line ($i$, $j$). For most scenarios in real power system (transmission network), the following engineering assumptions are often made:

1) The bus voltage magnitudes are approximately 1 p.u., which means $V_{i}\approx 1$ for all $i$.

2) The absolute value of phase angle differences across branches rarely exceeds $30^{\circ}$, hence,  $(\theta _{i}-\theta _{j})$ becomes small.

3) Let $r_{ij}$ and $x_{ij}$ denote the resistance and reactance of line ($i$, $j$), respectively. Typically, $r_{ij}$ is much smaller than  $x_{ij}$ for large-scale systems, that is, $r_{ij}<<x_{ij}$.

\subsection{Method 1: DC power flow model}

The DC power flow model is the most commonly used model due to its simplicity. The assumptions of the DC power flow model are: 
\begin{align}
   &V_{i}\approx 1, \textup{  }g_{ij}\approx 0,\textup{  }-b_{ij}\approx \frac{1}{x_{ij}}\\
   &\cos (\theta _{i}-\theta _{j})\approx 1,
   \textup{  }\sin (\theta _{i}-\theta _{j})\approx \theta _{i}-\theta _{j}
   \label{ass_voltage_angle}
\end{align}

\noindent Then, (\ref{acpf}) becomes:
\begin{align}
    P_{ij}=\frac{\theta _{i}-\theta _{j}}{x_{ij}}
\end{align}

\subsection{Method 2: using first-order Taylor series}

A linearization of the line flow equations is proposed in \cite{gvbs} based on first-order Taylor series: 
\begin{align}
   &V_{i}V_{j}\approx 1+(V_{i}-1)+(V_{j}-1)=V_{i}+V_{j}-1 \nonumber\\
   &V_{i}V_{j}\sin (\theta _{i}-\theta _{j})\approx \theta _{i}-\theta _{j}
\end{align}

Combined with the assumption of (\ref{ass_voltage_angle}), the linearized active power is obtained:
\begin{align}
    P_{ij}=g_{ij}(V_{i}-V_{j})-b_{ij}(\theta _{i}-\theta _{j})
\end{align}

\begin{table*}[]
\caption{Error of linear approximation for power flow}
\begin{center}
\label{table:lin_app}
\begin{tabular}{|c|c|c|c|c|c|}
\hline
\textbf{Test System}   & \textbf{Method 1}     & \textbf{Method 2}   & \textbf{Method 3} & \textbf{Method 4}    & \textbf{Method 5} \\ \hline
\textbf{14 bus system}            & 0.0647      & 0.0050    & 0.0197          & 0.0041     & 0.0015  \\ \hline
\textbf{57 bus system}            & 0.2678      & 0.0013    & 0.1400          & 0.0011     & 0.0006  \\ \hline
\textbf{200 bus system}  & 0.0538      & 0.0230    & 0.0062          & 0.0228     & 0.0059  \\ \hline
\textbf{500 bus system}  & 0.0638      & 0.0232    & 1.2628          & 0.0225     & 0.0060  \\ \hline
\textbf{2000 bus system} & 546627.7069 & 4021.1514 & 4481439588.9154 & 17970.2844 & 2.8350  \\ \hline
\end{tabular}
\end{center}
\end{table*}
\vspace{-0.1in}
\subsection{Method 3: modified phase angle model}

The model in \cite{0.95} introduces a constant adjustment factor to minimize the estimation errors in the assumption:
\begin{align}
   &\sin (\theta _{i}-\theta _{j})\approx 0.95 (\theta _{i}-\theta _{j})\nonumber\\
   &\cos (\theta _{i}-\theta _{j})\approx 0.95
\end{align}

\noindent And by assuming $V_{i}V_{j}\approx V_{i}^{2}\textup{ or }V_{j}^{2} $, the power flow becomes:
\begin{align}
    P_{ij}=0.95(g_{ij}(V_{i}^{2}-V_{j}^{2})-b_{ij}(\theta _{i}V_{i}^{2}-\theta _{j}V_{j}^{2}))\label{0.95pf}
\end{align} 

Here, $\theta _{i}V_{i}^{2}$ is the modified phase angle. It is regarded as an independent variable in optimization. 
\vspace{-0.15in}
\subsection{Method 4: using the square of voltage}
Z. Yang also proposed a linearized OPF model in \cite{0.5_2} which has a similar expression with (\ref{0.95pf}). Apart from the assumptions of voltage angle as (\ref{ass_voltage_angle}), the other assumptions of this model are:
\begin{align}
    V_{i}V_{j}(\theta _{i}-\theta _{j})\approx \theta _{i}-\theta _{j},~ V_{i}V_{j}(\theta _{i}-\theta _{j})^{2}\approx(\theta _{i}-\theta _{j})^{2}
\end{align}

\noindent Then we have:
\begin{align}
    P_{ij}=&g_{ij}\frac{V_{i}^{2}-V_{j}^{2}}{2}-b_{ij}(\theta _{i}-\theta _{j})\nonumber\\
   +&g_{ij}\left ( \frac{(V_{i}-V_{j})^{2}}{2}+ \frac{(\theta _{i}-\theta _{j})^{2}}{2}\right )
\end{align}

\noindent The linearized power flow equation then becomes:
\begin{align}
   & P_{ij}=g_{ij}\frac{V_{i}^{2}-V_{j}^{2}}{2}-b_{ij}(\theta _{i}-\theta _{j})
\end{align}

\noindent Here $V_{i}^{2}$ is regarded as an independent variable.
\vspace{-0.1in}
\subsection{Method 5: modiﬁed voltage magnitude method}
The model in \cite{log} uses the logarithmic transform of voltage magnitudes as the modiﬁed voltage magnitude:
\begin{align}
    U_{i}=\ln{V_{i}}
\end{align}

\noindent Based on the assumption that:
\begin{align}
  e^{-U^{i}}\approx 1-U^{i},\textup{  }e^{U^{j}+\textup{j}(\theta _{i}-\theta _{j})}\approx 1+U^{j}+\textup{j}(\theta _{i}-\theta _{j})  
\end{align}
 we obtain the following power flow:
\begin{align}
    P_{ij}(1-U_{i})=g_{ij}(U_{i}-U_{j})-b_{ij}(\theta _{i}-\theta _{j})\label{logpf}
\end{align}

Although there is a product term $P_{ij}(1-U_{i})$ in (\ref{logpf}), the model is fully linearized in OPF problems: 1) for PQ buses, $\sum_{j\in i}P_{ij}$ is known; 2) for PV buses, $U_{i}$ is known. Therefore, there is always only one variable in \ref{logpf}.

\subsection{Method 6: quadratic form of line loss}
\begin{figure}[htbp]
\centerline{\includegraphics[width=0.55\columnwidth]{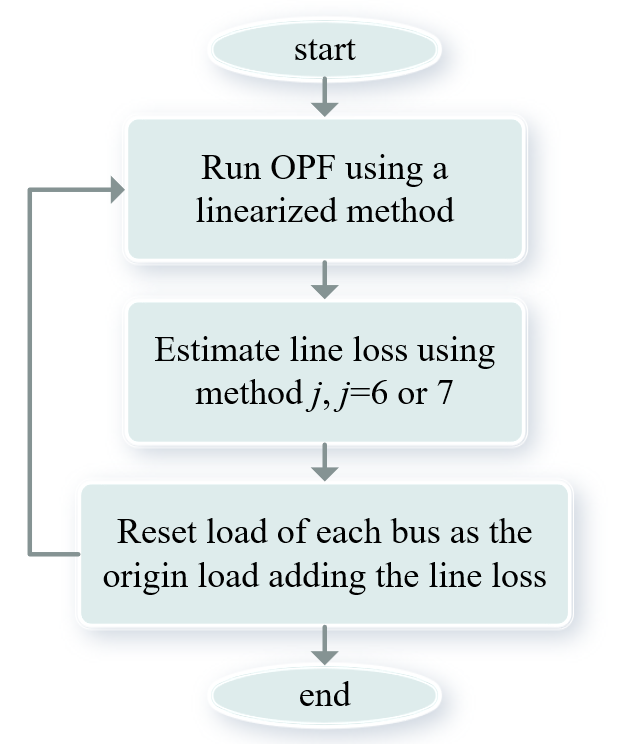}}
\caption{Iterative method considering line loss.}
\label{iteration}
\end{figure}

Consideration of line losses in optimal power flow problems is important to obtain a more realistic dispatch. One method to avoid implementation of the nonlinear line loss term into the optimal power flow problems is to perform an iterative approach, as shown in Fig.\ref{iteration}. This method is usually very fast and converges in a few iterations \cite{iteration}. The loss for line ($i$, $j$) can be written as:
\begin{align}
    P_{ij}^{loss}=g_{ij}\left ( V_{i}^{2}+V_{j}^{2} -2 V_{i}V_{j}\cos(\theta _{i}-\theta _{j})\right )\label{qua}
\end{align}

Reference \cite{0.5_2} made assumptions that
 $V_{i}\approx 1$ and $\cos(\theta _{i}-\theta _{j})=1-\frac{(\theta _{i}-\theta _{j})^{2}}{2}$. This assumption changes \eqref{qua} to,
\begin{align}
    P_{ij}^{loss}=g_{ij}(\theta _{i}-\theta _{j})^{2}
\end{align}
\subsection{Method 7: Proportional to power flow}
Another iterative method incorporating line loss is based on the AC model as in \cite{review2}:
\begin{align}
 P_{ij}^{loss}=I_{ij}^{2}r_{ij}=S_{ij}^{2}r_{ij}\approx(\alpha_{ij} P_{ij})^{2}r_{ij}   
\end{align}

\noindent where $I_{ij}$ and $S_{ij}$ are the current and the apparent power
of line ($i$, $j$), respectively. Here, $\alpha_{ij}$ is a scalar.

\begin{table*}[htbp!]
\caption{Optimality differences across methods.}\vspace{-4mm}
\label{table:optimality}
\begin{center}
\begin{tabular}{|c|c|c|c|c|c|c|}
\hline
      \textbf{Method}                    &  \textbf{Metric}    & \textbf{14 bus }     & \textbf{57 bus} & \textbf{200 bus} & \textbf{500 bus} & \textbf{2000 bus} \\ \hline
\multirow{3}{*}{Method 1} & $\varepsilon_{f} $  & 0.0119     & 0.0055 & 0.0004           & 0.0026           & 0.0571            \\ \cline{2-7} 
                          & $\varepsilon_{Pg} $ & 0.7802     & 0.1005 & 0.0005           & 0.0600           & 0.3618            \\ \cline{2-7} 
                          & $\varepsilon_{V} $  & 0.0113     & 0.0188 & 0.0480           & 0.0584           & 0.0624            \\ \hline
\multirow{3}{*}{Method 2} & $\varepsilon_{f} $  & 0.1891     & 0.0145 & 0.0697           & 0.9337           & 0.1203            \\ \cline{2-7} 
                          & $\varepsilon_{Pg} $ & 26808.7214 & 0.4784 & 0.4221           & 0.8651           & 6.5138            \\ \cline{2-7} 
                          & $\varepsilon_{V} $  & 0.0134     & 0.0192 & 0.0479           & 0.0576           & 0.0667            \\ \hline
\multirow{3}{*}{Method 3} & $\varepsilon_{f} $  & 0.1891     & 0.0145 & 0.0818           & 0.9283           & 0.1168            \\ \cline{2-7} 
                          & $\varepsilon_{Pg} $ & 26808.7214 & 0.4784 & 0.6847           & 0.8613           & 5.6751            \\ \cline{2-7} 
                          & $\varepsilon_{V} $  & 0.0134     & 0.0192 & 0.0479           & 0.0569           & 0.0646            \\ \hline
\multirow{3}{*}{Method 4} & $\varepsilon_{f} $  & 0.1891     & 0.0145 & 0.0817           & 0.9092           & 0.1177            \\ \cline{2-7} 
                          & $\varepsilon_{Pg} $ & 26808.7214 & 0.4784 & 0.7852           & 1.0317           & 5.9042            \\ \cline{2-7} 
                          & $\varepsilon_{V} $  & 0.0134     & 0.0192 & 0.0479           & 0.0575           & 0.0653            \\ \hline
\multirow{3}{*}{Method 5} & $\varepsilon_{f} $  & 0.0719     & 3.9516 & 0.0961           & 0.9961           & 0.2915            \\ \cline{2-7} 
                          & $\varepsilon_{Pg} $ & 1587.4024  & 3.7306 & 0.1153           & 1.0000           & 7.6650            \\ \cline{2-7} 
                          & $\varepsilon_{V} $  & 0.0115     & 0.0381 & 0.0480           & 0.0580           & 0.0868            \\ \hline
\multirow{3}{*}{Method 6} & $\varepsilon_{f} $  & 0.0115     & 0.0052 & 0.0004           & 0.0032           & 0.0006            \\ \cline{2-7} 
                          & $\varepsilon_{Pg} $ & 0.7802     & 0.0899 & 0.0005           & 0.0956           & 0.3336            \\ \cline{2-7} 
                          & $\varepsilon_{V} $  & 0.0113     & 0.0188 & 0.0480           & 0.0589           & 0.0624            \\ \hline
\multirow{3}{*}{Method 7} & $\varepsilon_{f} $  & 0.0115     & 0.0051 & 0.0004           & 0.0032           & 0.0011            \\ \cline{2-7} 
                          & $\varepsilon_{Pg} $ & 0.7799     & 0.0887 & 0.0005           & 0.0959           & 0.3332            \\ \cline{2-7} 
                          & $\varepsilon_{V} $  & 0.0113     & 0.0188 & 0.0480           & 0.0589           & 0.0624            \\ \hline
\end{tabular}
\end{center}
\end{table*}

\section{Comparison of different methods}
We compare the performances of the seven methods stated above in this section. We use the optimal results given in MATPOWER as the baseline and ignore the constraints for reactive power. The optimal results are denoted by *. We test different methods on 14 bus (i.e., case14), 57 bus (i.e., case57), 200 bus (i.e., case\_ACTIVSg200), 500 bus (i.e., case\_ACTIVSg500), and 2000 bus (i.e., case\_ACTIVSg2000) in MATPOWER. 
\vspace{-0.1in}
\subsection{Linear Approximation Error} 
The comparison of the accuracy of the considered linear approximations is shown in Table \ref{table:lin_app}. The average error of a particular linear approximation is given as:
\begin{align}
  \varepsilon = \sqrt{\frac{1}{N_r}\left ( \frac{P_{ij}^{[k]}(V^{*},\theta ^{*})-P_{ij}^{*}-\delta}{P_{ij}^{*}+\delta} \right ) ^{2}}  
\end{align}
Here, $N_r$ is the number of lines. $P_{ij}(V^{*},\theta ^{*})$ is obtained by applying different methods, $k=1,...5$ is the index of the method.  $P_{ij}^{*}$ is the result given in MATPOWER or equation (\ref{acpf}). $\delta $ is a small number added to avoid a zero denominator, and $\delta=10^{-7}$ in this paper. Since Method 6 and Method 7 are iterative methods, we don't test their performances in linear approximation. Method 1, the simplest DC model, performs the worst in terms of approximation error for most considered systems. This is understandable since the DC model is based on many assumptions. Further, we see that Method 5 is overall the most accurate linear approximation and is the most robust method. Method 5 even performs well in the 2000 bus system where all the other methods perform very badly.

\subsection{Optimality}
Next, we test the optimality of these methods. Here, Method 6 and Method 7 are based on Method 1, using four iterations. The models are tested as follows: First, we produce the output of each generator according to the linearized method. Then, we run an AC power flow using the chosen output of generators to obtain a steady-state operation. The voltage limit is ignored when obtaining the steady state. We compare this steady state with the baseline. The results are shown in Table \ref{table:optimality}.  We define the error of the objective function, generator outputs, and voltage as (\ref{accuracy1}), (\ref{accuracy2}), and (\ref{accuracy3}). Here, we use superscript $k$ to denote the results obtained by method $k$. $N_g$ and $N_b$ are the number of generators and buses, respectively.

\begin{align}
\label{accuracy1}
& \varepsilon_{f}=\left | \frac{f^k-(f^*+\delta)}{f^*+\delta} \right |\\
\label{accuracy2}
 & \varepsilon_{Pg}=\sqrt{\frac{1}{N_g}\sum _{i=1}^{N_g}\left (\frac{P_g^{i[k]}-\left (P_g^{i*}+\delta  \right )}{ (P_g^{i*}+\delta  )} \right )^{2}}
 \end{align}
\begin{align}
 \label{accuracy3}
 & \varepsilon_{V}=\sqrt{\frac{1}{N_b}\sum _{i=1}^{N_b}\left (\frac{V_i^{[k]}-V_i^{*}}{V_i^{*}} \right )^{2}}
\end{align}
\begin{table}[h!]
\caption{results of feasibility}\vspace{-4mm}
\label{table:feasibility}
\begin{center}
\begin{tabular}{|c|c|c|c|c|}
\hline
\textbf{Method}                   & \textbf{Metric}               & \textbf{14 bus} & \textbf{57 bus} & \textbf{2000 bus} \\ \hline
\multirow{4}{*}{Method 1} & $\frac{N_{out }}{N_b}$   & 0.2143 & 0.0175 & 0.0020            \\ \cline{2-5} 
                          & $N_{above} $             & 0      & 1      & 0                 \\ \cline{2-5} 
                          & $N_{below} $             & 3      & 0      & 4                 \\ \cline{2-5} 
                          & $\varepsilon_{V}^{out} $ & 0.0189 & 0.0151 & 0.0286            \\ \hline
\multirow{4}{*}{Method 2} & $\frac{N_{out }}{N_b}$   & 0.2143 & 0.0175 & 0.0025            \\ \cline{2-5} 
                          & $N_{above} $             & 0      & 1      & 5                 \\ \cline{2-5} 
                          & $N_{below} $             & 3      & 0      & 0                 \\ \cline{2-5} 
                          & $\varepsilon_{V}^{out} $ & 0.0187 & 0.0162 & 0.2479            \\ \hline
\multirow{4}{*}{Method 3} & $\frac{N_{out }}{N_b}$   & 0.2143 & 0.0175 & 0.0025            \\ \cline{2-5} 
                          & $N_{above} $             & 0      & 1      & 5                 \\ \cline{2-5} 
                          & $N_{below} $             & 3      & 0      & 0                 \\ \cline{2-5} 
                          & $\varepsilon_{V}^{out} $ & 0.0187 & 0.0162 & 0.1980            \\ \hline
\multirow{4}{*}{Method 4} & $\frac{N_{out }}{N_b}$   & 0.2143 & 0.0175 & 0.0025            \\ \cline{2-5} 
                          & $N_{above} $             & 0      & 1      & 5                 \\ \cline{2-5} 
                          & $N_{below} $             & 3      & 0      & 0                 \\ \cline{2-5} 
                          & $\varepsilon_{V}^{out} $ & 0.0187 & 0.0162 & 0.2056            \\ \hline
\multirow{4}{*}{Method 5} & $\frac{N_{out }}{N_b}$   & 0.2143 & 0.1930 & 0.0235            \\ \cline{2-5} 
                          & $N_{above} $             & 0      & 11     & 43                \\ \cline{2-5} 
                          & $N_{below} $             & 3      & 0      & 4                 \\ \cline{2-5} 
                          & $\varepsilon_{V}^{out} $ & 0.0191 & 0.0479 & 0.3392            \\ \hline
\multirow{4}{*}{Method 6} & $\frac{N_{out }}{N_b}$   & 0.2143 & 0.0175 & 0.0000            \\ \cline{2-5} 
                          & $N_{above} $             & 0      & 1      & 0                 \\ \cline{2-5} 
                          & $N_{below} $             & 3      & 0      & 0                 \\ \cline{2-5} 
                          & $\varepsilon_{V}^{out} $ & 0.0189 & 0.0152 & 0.0000            \\ \hline
\multirow{4}{*}{Method 7} & $\frac{N_{out }}{N_b}$   & 0.2143 & 0.0175 & 0.0000            \\ \cline{2-5} 
                          & $N_{above} $             & 0      & 1      & 0                 \\ \cline{2-5} 
                          & $N_{below} $             & 3      & 0      & 0                 \\ \cline{2-5} 
                          & $\varepsilon_{V}^{out} $ & 0.0189 & 0.0152 & 0.0000            \\ \hline
\end{tabular}
\end{center}
\end{table}

Method 2-4 usually produce more optimal results when including reactive power, according to \cite{review1}, \cite{review2}. However, we see that Method 2-5 are not robust enough compared to DC-based methods (Method 1, Method 6, and Method 7) when we only consider active power. Method 6 and Method 7 generally perform better than Method 1. This is likely because including the line loss helps to obtain a more accurate objective function. However, they do not improve the voltage accuracy.

\begin{table*}[h!]
\caption{Execution time (s)}\vspace{-4mm}
\label{table:time}
\begin{center}
\begin{tabular}{|c|c|c|c|c|c|c|c|}
\hline
\textbf{Test system}    & \multicolumn{1}{c|}{\textbf{Method 1}} & \multicolumn{1}{c|}{\textbf{Method 2}} & \multicolumn{1}{c|}{\textbf{Method 3}} & \multicolumn{1}{c|}{\textbf{Method 4}} & \multicolumn{1}{c|}{\textbf{Method 5}} & \multicolumn{1}{c|}{\textbf{Method 6}} & \multicolumn{1}{c|}{\textbf{Method 7}} \\ \hline
\textbf{14 bus system}            & 0.91                          & 2.63                          & 2.57                          & 2.53                          & 2.76                          & 5.34                          & 5.25                          \\ \hline
\textbf{57 bus system}            & 1.56                          & 3.70                          & 3.71                          & 3.67                          & 4.61                          & 7.01                          & 6.93                          \\ \hline
\textbf{200 bus system}  & 2.61                          & 9.81                          & 10.06                         & 9.91                          & 10.20                         & 11.47                         & 11.66                         \\ \hline
\textbf{500 bus system}  & 5.04                          & 34.16                         & 34.43                         & 31.56                         & 33.12                         & 20.69                         & 20.10                         \\ \hline
\textbf{2000 bus system} & 13.99                         & 1008.59                       & 1005.85                       & 1011.97                       & 1003.54                       & 59.47                         & 59.00                         \\ \hline
\end{tabular}
\end{center}
\end{table*}

\subsection{Feasibility}
Here we test the feasibility of these methods by observing how many buses violate the voltage limits using the linearization. We adopt the same testing process as that for optimality: compare the steady state with the baseline. Since no bus violates the voltage limit in the 200 bus system and 500 bus system, their results are not shown. Here, $N_{out}$ is the number of buses with voltage violations, and $N_{above},N_{below}$ give the number of buses above and below the limit respectively. $\varepsilon_{V}^{out} $ is the average voltage error of these buses:
\begin{align}
\varepsilon_{V}^{out}=\sqrt{\frac{1}{N_{out }}\sum _{i=1}^{N_{out }}\left (\frac{V_i^{[k]}-V_i^{*}}{V_i^{*}} \right )^{2}}  
\end{align}

The feasibility results are shown in Table \ref{table:feasibility}. We could see that Method 2-5 generally worsen the voltage issue compared to DC-based methods.  Method 5 performs the worst, causing many buses to have voltage violations, and Method 6 and Method 7 perform better because they help decrease the number of buses with voltage issues.  

\subsection{Execution Time}
The execution time of running each method 100 times is shown in Table \ref{table:time}. Method 1 is the fastest. Since Method 6 or Method 7 can be regarded as iterative progress based on Method 1, the time for Method 6 or Method 7 is slightly over four times as that for Method 1. When the system is relatively small, Method 2-5 need about twice the running time of Method 1. However, as the network size grows, the time for Method 2-5 grows faster. The running time for the 2000-buses system using Method 2-5 is 70 times longer than that using Method 1.

\section{Strengths and weaknesses analysis}
\begin{figure}[htbp]
\centerline{\includegraphics[trim={0 1.5cm 0 3.6cm},width=0.9\columnwidth]{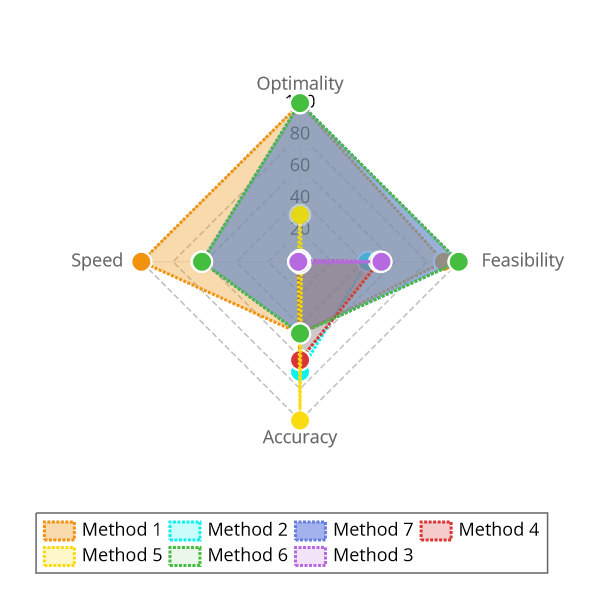}}
\caption{Advantages of different methods.}
\label{figure:radar}
\end{figure}

We rate different methods according to their performances in accuracy, optimality, feasibility, and speed as described below. \textbf{Accuracy} is calculated by summing up the errors across all cases in Table \ref{table:lin_app}. Since Method 6-7 do not have these log values, we set them the same as that of Method 1. \textbf{Optimality} is calculated by summing the errors of all variables in Table \ref{table:optimality}. \textbf{Feasibility} is calculated by summing $\frac{N_{out }}{N_b}$ and $\varepsilon_{V}^{out}$ in Table \ref{table:feasibility}. \textbf{Speed} is calculated by summing the running times in Table \ref{table:time}. For each of these metrics, lower values are better. In order to simplify the analysis, we transform these values by applying the log function to their reciprocal value (such that higher numbers are better). Then, we transform these log values linearly to 1-100. Fig. \ref{figure:radar} presents the radar chart of these methods. We could see that Method 1 covers the largest area, hence, it is the most robust method.

\section{Conclusion}
This paper compares and contrasts seven linear approximations for active power in OPF problems. Apart from accuracy, optimality, and computational time, this paper also considers the feasibility of different models. The results show that: (1) Method 5 using the logarithmic transform of voltage magnitudes achieves the highest accuracy of linear approximation. (2) DC power flow has the smallest error when ignoring reactive power in optimal power flow problems. The iterative methods considering line loss could further decrease errors and obtain a more accurate optimal solution. (3) The iterative methods based on DC power flow could also decrease the number of buses with voltage violations. (4) DC power flow is the fastest method and is generally the most robust.  

\vspace{12pt}


\begin{thebibliography}{00}
\bibitem{intro1}CAISO, "Market processes and products",California ISO - Market Processes, http://www.caiso.com/market/Pages/MarketProcesses.aspx
\bibitem{intro2}M. Li, W. Huang, N. Tai, L. Yang, D. Duan and Z. Ma, "A Dual-Adaptivity Inertia Control Strategy for Virtual Synchronous Generator," in IEEE Trans. on Power Sys., vol. 35, no. 1, pp. 594-604, Jan. 2020.
\bibitem{review2}D. Yu, J. Cao and X. Li, "Review of power system linearization methods and a decoupled linear equivalent power flow model," 2018 International Conference on Electronics Technology (ICET), 2018, pp. 232-239.
\bibitem{intro6}H. Wang, Y. Wang, C. Gao, L. Xu, J. Hou and X. Tao, "Iterative algorithm of DC power flow based on network loss equivalent load model", Automation of Elec. Power Sys., vol. 39, pp. 99-103, Jan. 2015.
\bibitem{intro7}J. Yang, N. Zhang, C. Kang and Q. Xia, "A State-Independent Linear Power Flow Model with Accurate Estimation of Voltage Magnitude", IEEE Trans. on Power Systems, vol. 32, pp. 3607-3617, Sept. 2017.
\bibitem{intro8}Y. Wang, N. Zhang, H. Li, J. Yang and C. Kang, "Linear three-phase power flow for unbalanced active distribution networks with PV nodes," in CSEE Journal of Power and Energy Systems, vol. 3, no. 3, pp. 321-324, Sept. 2017.
\bibitem{BakerACM}K. Baker, ``Solutions of DC OPF Are Never AC Feasible,'' Association for Computing Machinery (ACM) e-Energy, 2021. 
\bibitem{feasibility1}Mingyang Li, Qianchuan Zhao and P. B. Luh, "DC power flow in systems with dynamic topology," 2008 IEEE Power and Energy Society General Meeting, 2008, pp. 1-8.
\bibitem{feasibility2}M. Li, Q. Zhao and P. B. Luh, "Decoupled load flow and its feasibility in systems with dynamic topology," 2009 IEEE Power Energy Society General Meeting, 2009, pp. 1-8.
\bibitem{review1}Yang, Z., Zhong, H., Xia, Q. and Kang, C. (2017), Solving OPF using linear approximations: fundamental analysis and numerical demonstration. IET Gener. Transm. Distrib., 11: 4115-4125.
\bibitem{gvbs} Zhang, H., Heydt, G.T., Vittal, V., et al.: ‘An improved network model for transmission expansion planning considering reactive power and network losses’, IEEE Transactions on Power Systems vol. 28, no. 3, pp. 3471–3479, 2013.
\bibitem{0.95} S. M. Fatemi, S. Abedi, G. B. Gharehpetian, S. H. Hosseinian and M. Abedi, "Introducing a Novel DC Power Flow Method With Reactive Power Considerations," in IEEE Transactions on Power Systems, vol. 30, no. 6, pp. 3012-3023, Nov. 2015.
\bibitem{0.5_2} Yang, Z., Zhong, H., Bose, A., et al.: ``A linearized OPF model with reactive power and voltage magnitude: a pathway to improve the Mw-only DC OPF,'' IEEE Transactions Power Systems vol. 33, no. 2, pp. 1734 - 1745, Mar. 2018.
\bibitem{log} Z. Li, J. Yu and Q. H. Wu, "Approximate Linear Power Flow Using Logarithmic Transform of Voltage Magnitudes With Reactive Power and Transmission Loss Consideration," in IEEE Transactions on Power Systems, vol. 33, no. 4, pp. 4593-4603, July 2018.
\bibitem{iteration}T. N. dos Santos and A. L. Diniz, "A Dynamic Piecewise Linear Model for DC Transmission Losses in Optimal Scheduling Problems," in IEEE Transactions on Power Systems, vol. 26, no. 2, pp. 508-519, May 2011.

\bibitem{kargarian2016toward} A. Kargarian, J. Mohammadi, J. Guo, S. Chakrabarti, M. Barati, G. Hug, S. Kar, and R. Baldick “\textit{Toward Distributed/D-Centralized Optimal Power Flow Implementation in Future Electric Power Systems}”, IEEE Transactions on Smart Grid, Vol. 9, Issue. 4, July 2018




\end{thebibliography}
\end{document}